\documentclass[10pt,preprint]{aastex}

\newcommand{\kms}{\,km\,s$^{-1}$}

\newcommand{\arcs}{$^{\prime\prime}$}

\newcommand{\eg}{$e. g.$}
\newcommand{\fuse}{$FUSE$}
\newcommand{\FUSE}{$FUSE$}

\slugcomment{Accepted for publication in the Astrophysical Journal}

\shorttitle{FUSE Survey of \ion{O}{6} in the SMC}
\shortauthors{Hoopes et al.}

\begin{document}

\title{A {\it FUSE} Survey of Interstellar \ion{O}{6} Absorption in the Small
Magellanic Cloud }

\author{Charles G. Hoopes\altaffilmark{1}, Kenneth R. Sembach\altaffilmark{1,}\altaffilmark{2}, J. Christopher Howk\altaffilmark{1}, Blair D. Savage\altaffilmark{3},\\ and Alex W. Fullerton\altaffilmark{1,}\altaffilmark{4}} 
\altaffiltext{1}{Department of Physics and Astronomy, Johns Hopkins University,
3400 N. Charles St., Baltimore, MD 21218; choopes@pha.jhu.edu, howk@pha.jhu.edu, awf@pha.jhu.edu}
\altaffiltext{2}{Current address: Space Telescope Science Institute, 3700 San Martin Dr., Baltimore, MD 21218; sembach@stsci.edu}
\altaffiltext{3}{Department of Astronomy, University of Wisconsin-Madison, 475
North Charter Street, Madison, WI 53706; savage@astro.wisc.edu}
\altaffiltext{4}{Department of Physics and Astronomy, University of Victoria, P.O. Box 3055, Victoria, BC V8W 3P6, Canada}

\begin{abstract}

We present the results of a {\it Far Ultraviolet Spectroscopic
Explorer (FUSE)} survey of \ion{O}{6} 1031.93~\AA\ and
1037.62~\AA\ absorption toward 18 OB stars in the Small Magellanic Cloud
(SMC). The \FUSE\ data are of very high quality, allowing a detailed
study of the coronal temperature gas in the SMC. We find that
\ion{O}{6} is ubiquitous in the SMC, with a detection along every sight
line. The average value of the \ion{O}{6} column density in the SMC is
log $<N$(\ion{O}{6})$>=14.53$. This value is $1.7$ times higher than
the average value for the Milky Way halo (perpendicular to the
Galactic plane) of log $N_\perp$(\ion{O}{6})=14.29 found by \fuse,
even though the SMC has much lower metallicity than the Galaxy. The
column density in the SMC is higher along sight lines that lie close
to star-forming regions, in particular NGC~346 in the northern part of
the SMC, and to a lesser degree the southwestern complex of \ion{H}{2}
regions. This correlation with star formation suggests that local
processes have an important effect on the distribution of coronal gas
in the SMC. If the sight lines within NGC~346 are excluded, the mean
column density for the SMC is log $N$(\ion{O}{6})$=14.45$, only $1.4$
times higher than the Milky Way average. The standard deviation of the
column densities for sight lines outside of NGC~346 is $\pm27$\%,
somewhat lower than the deviation seen in the Milky Way halo. The
lowest \ion{O}{6} column densities, log $N$(\ion{O}{6})$\sim14.3$,
occur in the central region and in the southeastern ``Wing'' of the
galaxy. Even these low column densities are as high as the Milky Way
average, establishing the presence of a substantial, extended component of
coronal gas in the SMC. The \ion{O}{6} absorption is always
shifted to higher velocities than the main component of lower
ionization gas traced by \ion{Fe}{2} absorption. The \ion{O}{6} line
widths are broader than expected for pure thermal broadening at
$3\times10^5$~K, the temperature at which the \ion{O}{6} peaks in
abundance, so large non-thermal motions or multiple hot gas components
are likely present. We discuss several mechanisms that may be able to
explain the observed properties of the hot gas, including supershells,
a galactic fountain, and the infall of gas previously stripped from
the SMC by tidal interactions with the Milky Way and the Large
Magellanic Cloud. If a galactic fountain produces the hot gas, the
mass flux per unit surface area is $\dot M/\Omega \sim
2\times10^{-2}$~M$_{\odot}$~yr$^{-1}$~kpc$^{-2}$.

\end{abstract}

%% Keywords should appear after the \end{abstract} command. The uncommented
%% example has been keyed in ApJ style. See the instructions to authors
%% for the journal to which you are submitting your paper to determine
%% what keyword punctuation is appropriate.


%% From the front matter, we move on to the body of the paper.
%% In the first two sections, notice the use of the natbib \citep
%% and \citet commands to identify citations.  The citations are
%% tied to the reference list via symbolic KEYs. The KEY corresponds
%% to the KEY in the \bibitem in the reference list below. We have
%% chosen the first three characters of the first author's name plus
%% the last two numeral of the year of publication as our KEY for
%% each reference.

\section{Introduction}
%section 1

The \ion{O}{6} 1031.93~\AA\ and 1037.62~\AA\ lines are important
diagnostics of the processes responsible for distributing mass and
energy in the interstellar medium (ISM). \ion{O}{6} is difficult to
produce through photoionization, requiring photons with $h\nu \ge 114$
eV. Gas containing \ion{O}{6} is therefore most likely collisionally
ionized.  Since \ion{O}{6} peaks in abundance at $3\times10^5$~K
\citep{sd93}, an unstable region of the cooling curve, gas containing
\ion{O}{6} is either cooling from higher temperatures, or being
heated. As such, it is a useful probe of energetic processes in the
ISM including shock heating \citep{sm79}, conductive heating and
cooling \citep{bbf90}, and turbulent mixing layers
\citep{ssb93}. Coronal gas traced by \ion{O}{6} is intermediate in
temperature between the very hot ($T\sim10^{6-7}$~K) X-ray emitting
gas and the warm ($T\sim10^4$~K) ionized medium.

Opportunities to observe the \ion{O}{6} lines have been limited. The
first observations of \ion{O}{6} were made with the {\it Copernicus}
satellite along sight lines toward nearby bright stars ($V\le7$,
d $\la2000$~pc) \citep{jm74,j78a,j78b}. Later observations were
possible with the ORFEUS-SPAS missions
\citep{hb96,h98,w98,ssh99}. These observations, combined with
observations of \ion{C}{4} $\lambda\lambda1548.20, 1550.77$,
\ion{N}{5} $\lambda\lambda1238.82, 1242.80$, and \ion{Si}{4}
$\lambda\lambda1393.76, 1402.77$, have built upon the original idea of
a Galactic corona of hot gas which provides pressure confinement for
high-latitude clouds \citep{s56}. While results for \ion{C}{4},
\ion{N}{5}, and \ion{Si}{4} have greatly increased our understanding
of the Galactic corona, the relative lack of information on \ion{O}{6}
has been frustrating, since it probes a hotter temperature range and
is less likely to be produced by photoionization.

With the launch of the {\it Far Ultraviolet Spectroscopic Explorer
(FUSE)} in 1999 \citep{m00}, access to the \ion{O}{6} lines has
returned. \FUSE\ is much more sensitive than {\it Copernicus}, allowing
more distant stars and extragalactic sources to be observed. One of
the major science goals of \FUSE\ is to study the properties and
distribution of hot gas in the local universe through
\ion{O}{6} absorption. Early results have already been published for
the Galactic halo and high velocity clouds \citep{s00,sem00}, and more
comprehensive surveys are in progress.

The sensitivity of \FUSE\ makes it possible to observe early type
stars in the Magellanic Clouds, so we can now study the \ion{O}{6}
absorption in galaxies with different properties than our own. The
Small Magellanic Cloud (SMC) has properties that are vastly different
from the Milky Way. It has low mass, low metallicity, and it is
currently undergoing a gravitational interaction with two more massive
galaxies. These factors may influence the properties of hot gas in the
ISM of the SMC. The hot gas in the SMC has been studied previously
using {\it International Ultraviolet Explorer (IUE)} spectra
\citep{ds80,fs83,f84,f85,fs85}.  Absorption by \ion{C}{4} and
\ion{Si}{4} was measured toward the nine stars studied by \cite{fs85},
who found evidence for a global component of highly ionized gas. More
recently, Space Telescope Imaging Spectrograph (STIS) spectra of one
of the sight lines studied by \cite{fs85}, AV229 (HD 5980), were
analyzed by \cite{k01} and \cite{hshb01}, who also found evidence for
hot gas in a supernova remnant in addition to the general ISM. The
spectra show absorption by \ion{N}{5} in addition to \ion{C}{4} and
\ion{Si}{4}. \fuse\ data on \ion{O}{6} toward Sk~108 in the SMC has
already been analyzed by \cite{m01}, but the complicated stellar
continuum prevented an accurate determination of the \ion{O}{6} column
density. \cite{w91} and \cite{ww92} discovered an X-ray halo around
the SMC using {\it Einstein} data, with spectral properties that
suggest the presence of hot, thermal ISM components.

In this paper we present the results of a \FUSE\ \ion{O}{6} absorption
survey of 18 stars in the SMC.  By observing stars in a variety of
environments in the SMC, we search for a global component of hot gas, as
well as clues to the origins of the highly ionized gas. In section 2
we describe the observations and data processing. Section 3 describes
the \ion{O}{6} content of the SMC, its variation with location, and a
comparison to the Milky Way. The kinematics of the \ion{O}{6}
absorbing gas are described in section 4, and section 5 contains a
discussion of the observed \ion{O}{6} absorption. Section 6 summarizes
the primary conclusions of our study. In a parallel effort,
\cite{hsfs02} present a study of \ion{O}{6} absorption in the Large
Magellanic Cloud.

\section{Observations and Data Reduction}
%section 2

We obtained FUSE spectra of early type stars in 18 locations in the
SMC.  Table 1 lists the names and properties of the stars used in this
project. For information on the local interstellar environment of
these stars, see \cite{dhfbs02}. The star names are from
\cite{av82} and \cite{sk68,sk69}. The NGC~346 identifiers are from \cite{w86},
\cite{w95}, and \cite{mpg89}, and the new spectral type for NGC346$-$3
is from \cite{w01a}. The parameters of the \FUSE\ observations are
given in Table 2. All of the spectra were taken through the large
(LWRS) apertures, which are 30\arcs\,$\times$\,30\arcs\ in size.  The
\FUSE\ instrument consists of four channels, two optimized for short
ultraviolet wavelengths (SiC1 and SiC2: 905--1100\,\AA) and two
optimized for longer ultraviolet wavelengths (LiF1 and LiF2:
1000--1187\,\AA).  The \ion{O}{6} lines are covered by all four
channels, but the LiF1 channel is the most sensitive; it is the
primary channel considered here, although all of the results were
verified with the LiF2 data. Because of fixed pattern features in the
detectors, we required that any absorption line be present in these
two channels to be considered a real feature. The LiF1 spectra in the
wavelength region containing the \ion{O}{6} lines are shown in Figure
1.

The raw time-tagged photon event lists for each individual exposure
were combined and run through the standard \FUSE\ calibration
pipeline (CALFUSE v1.8.7) available at Johns Hopkins University. The
pipeline screens data for passage through the South Atlantic Anomaly
and low Earth limb angle pointings and corrects for thermal drift of
the gratings, thermally induced changes in the detector read-out
circuitry, and Doppler shifts due to the orbital motion of the
satellite. Finally the pipeline subtracts a constant detector background and
applies wavelength and flux calibration. The velocity zero-point was
set by shifting the spectrum so that the Milky Way component of the
H$_2$ and low ionization lines lie at a Local Standard of Rest (LSR)
velocity of 0\,\kms, based on the results of \cite{m01}. The spectra
were binned by three pixels ($\sim$6\,\kms\ near \ion{O}{6}), and the
nominal resolution of the data is $\sim$20\,\kms\ (FWHM). The
wavelength solution is accurate to $\sim$6\,\kms\ on average, although
deviations of 10--15\,\kms\ may exist over small regions of the
spectrum.

The four sight lines in NGC~346 present an additional
complication. The four stars \FUSE\ observed are separated by only a
few arcseconds, so the light of all four stars, and perhaps others,
entered the LWRS apertures and is present in the spectrum of each
star. The resultant spectra are blends of several stars, with
broadened absorption features due to the positional offsets of the
sources in the apertures. The four spectra in this region look very
similar, with minor differences probably caused by shifting a small
number of fainter stars in or out of the aperture at the different
pointings. Although we analyzed each sight line separately, the reader
should recognize that the results for these four sight lines are
correlated. When determining global properties of the SMC or Milky
Way, we averaged the results for these four sight lines, and treated
them as one measurement. In a future paper we will report on new
\fuse\ observations of the NGC~346 stars using smaller apertures to
isolate the individual stars.

\section{Column Densities}
%section 3

\subsection{Measurements}

Equivalents widths and column densities were measured using the
procedures described by \cite{ss92}. The total column densities were
derived by summing the apparent optical depth over the velocity range
of the \ion{O}{6} 1031.93~\AA\ line. The \ion{O}{6} 1037.62~\AA\ line
was blended with other lines in all of the spectra, rendering it
useless for determining \ion{O}{6} column densities. The apparent
optical depth method is valid as long as there is no unresolved
saturated structure in the line. The high temperatures required for
\ion{O}{6} would result in a thermal line width of $\approx 30$~\kms
(FWHM), larger than the \FUSE\ line spread function width ($\approx
20$~\kms), so the \ion{O}{6} lines should be resolved. No saturated
\ion{O}{6} lines were seen in the \FUSE\ data, so we conclude that the
absorption is optically thin.

Uncertainties in the equivalent widths and column densities were
evaluated very conservatively. For each spectrum we first placed a
nominal continuum over the \ion{O}{6} 1031.93~\AA\ line by fitting low
order ($\le5$) Legendre polynomials to the nearby stellar
continuum. Continuum error estimates were calculated according to the
prescription outlined by \cite{ss92}.  However, due the
high degree of continuum curvature and systematic uncertainties caused
by possible stellar absorption near some of the profiles, we explored
a wider range of parameter space in the continuum fits than was
suggested by the formal (objective) approach. To do this we varied the
continuum fit to the most extreme values above and below the nominal
fit that were still reasonable continuations of the nearby
continuum. These bracketing continuum placements are shown in Figure 1
as dashed lines, while the nominal fits are shown as solid
lines. While this method adds a more subjective element to the
evaluation of uncertainties, the potentially drastic effect of errors
in the continuum placement justify such a conservative approach.

We used the 6-0~P(3) 1031.20~\AA\ and 6-0~R(4) 1032.35~\AA\ lines
of H$_2$, which are close to the \ion{O}{6} line, to test whether the
continuum fits were reasonable. To do this we fit a Gaussian function
to nearby H$_2$ lines from the same rotational level having similar
values of $f\lambda$, and plotted the fits at the positions of the
H$_2$ lines near \ion{O}{6}. The fits are plotted in Figure 2, which
show the continuum-normalized \ion{O}{6} lines using the nominal
continuum fits. In most cases the fits matched the observed lines
using the nominal continuum fit. In some of the spectra there is a
stellar feature that falls on the 6-0~P(3) 1031.195~\AA\ line (\eg,
AV378). This feature appears to be correlated in strength with another
feature at $\sim1033.5$~\AA. Other cases where the H$_2$ line fit did
not match the data are discussed below. All \ion{O}{6} measurements
were carried out using the three continuum fits in order to evaluate
the uncertainties resulting from errors in the continuum
placement. The presence of variable features in the stellar winds of
the stars can cause substantial errors in the derived \ion{O}{6}
column densities by affecting the continuum placement \citep{l01}. We
have not made any attempt to correct for this directly, but the sample
was defined by choosing stars that appeared to have relatively simple
continua around the 1032~\AA\ line.

The H$_2$ model for AV69 did not match the 6-0~R(4) 1032.351~\AA\
line. This star has a peculiar metallicity, which can affect the shape
of the stellar wind lines \citep{w00}. To test the hypothesis that the
peculiar wind lines have compromised our continuum fit, we have
examined the STIS spectrum of AV69, along with eight other stars in
the \fuse\ sample that have been observed by STIS \citep{w00}. While
the \ion{C}{4} lines show a normal P-Cygni wind profile, the
\ion{N}{5} lines have very weak and relatively narrow wind profiles
\citep{w00}. If the \ion{O}{6} stellar wind profile is shaped
similarly, it would hinder the measurement of the interstellar
\ion{O}{6}. This would explain the lack of agreement between the H$_2$
model and the observed spectrum. These tests lead us to distrust the
interstellar \ion{O}{6} column density measured toward AV69, which
would be the largest in the SMC if the continuum fit were correct. The
H$_2$ models for AV14 and AV26 also did not match the observed spectra
well. Unfortunately these two stars have not been observed by STIS, so
we cannot be sure if there is contamination by stellar
absorption. Absorption by \ion{O}{6} in the SMC at high velocities
could also produce this effect. Nevertheless, these sight lines are
suspicious, and while the measurements are included with those of the
other sight lines, the reader is warned to interpret the results for
these three stars with caution. For global properties we computed
averages both with and without AV14 and AV26. AV69 is not included in
the global averages as it is clearly not reliable.

A nominal velocity integration range was chosen for the Milky Way and
SMC components of \ion{O}{6} in each spectrum. We chose the velocity
of the point of lowest apparent optical depth between the two
components to be the boundary for the integration limits. If
the two components are described by overlapping Gaussian functions,
the approach could shift some of the column density from the stronger
component to the weaker component. Tests using Gaussian fits to
several of the sight lines revealed that this effect is $\la$3\% in the
worst cases, and $\la$1\% in typical cases. To account for this
effect, as well as possible H$_2$ absorption on either side of the
\ion{O}{6} 1032.93~\AA\ line, we varied the velocity integration
limits by $\pm5$~\kms\ so that the total integration range was
10~\kms\ larger or smaller than the nominal limits. The variation in
column density found with this method was greater than the effect of the
overlapping Gaussian profiles. This uncertainty was combined with the
fixed-pattern noise uncertainty in the \FUSE\ spectra. The measured
equivalent widths and column densities for the Milky Way and the SMC
components are listed in Tables 3 and 4. The first quoted uncertainty
is that due to fixed pattern noise and velocity range errors, and the
second shows the effects of varying the continuum placement (as
indicated in Figure 1), which in the majority of cases is the dominant
source of uncertainty.

\subsection{Results}

It is immediately clear from the spectra shown in Figures 1 and 2 that
the \ion{O}{6} absorption from the SMC is usually much stronger than
that in the Milky Way along the same line of sight. \ion{O}{6} is
detected in both the Milky Way and the SMC along every SMC sight line
observed by \fuse, including the sight line toward Sk~188, which is in
the periphery of the SMC. This suggests that a widespread component of coronal
temperature gas is present in the SMC. The measured equivalent widths,
column densities, line widths, and positions are given in Tables 3 and
4 for the Galactic and SMC absorption, respectively.

Figure 3 shows the locations of the \fuse\ targets on an H$\alpha$
image of the SMC \citep{g01}. The stars are marked with circles denoting the
\ion{O}{6} column density, and are labeled by the identification number
from Table 1. The largest star-forming region in the SMC is NGC~346,
in the northern part of the galaxy at the location of star 8 in Figure
3. Four of the targets are in this \ion{H}{2} region, and several
others lie close to it. There is another concentration of
\ion{H}{2} regions in the southwest corner of the SMC, also sampled by
several \FUSE\ sight lines. The central and the southeastern parts of the
galaxy are much less actively forming stars. The southeastern
extension is sometimes called the ``Wing'', while the northern and
southwestern parts make up the ``Bar.'' Most of the star formation in
the SMC is occurring in the Bar \citep{k95}.

The largest \ion{O}{6} column densities occur near the star-forming
regions. Figure 4 shows the distribution of log $N$(\ion{O}{6}) with
distance from NGC~346. The column toward AV229 (Sk~78, HD5980) is the
highest (excluding AV69, which may be compromised by stellar
absorption), with  log $N$(\ion{O}{6})=$14.86$. The sight line to AV229 passes through a
supernova remnant (SNR) in the SMC \citep{k01}, which contributes
roughly one third of the total \ion{O}{6} column \citep{hshb01}. Even
if the contribution of the SNR is excluded, the \ion{O}{6} column is
one of the highest in the SMC, with log $N$(\ion{O}{6})=14.68. AV229
is on the edge of NGC~346, and high columns are seen toward stars in
and around this \ion{H}{2} region (AV232 and the NGC~346 stars). The
stars outside of NGC~346 in the northern end of the Bar are markedly
lower, however. The \ion{O}{6} column toward AV321 (star 12) is still
relatively high, but the value for AV378 (star 13) is low compared to
most of the other stars in the sample.

The highest points outside of NGC~346 are the points in the
star-forming regions in the southwestern end of the Bar, marked by
open circles in Figure 4. The three highest values in this direction
are found for AV69, AV14, and AV26, all of which may be contaminated
by stellar absorption and are not shown in Figure 4. If these are left
out, three of the remaining sight lines in this area (AV95, AV75, and
AV15, stars 7, 5, and 2) still show large \ion{O}{6} columns, but not
as large as those near NGC~346, and one star (AV83, star 6) is fairly
low.  The \ion{H}{2} regions in this part of the SMC are not forming
stars as actively NGC~346, based on their H$\alpha$ luminosity
\citep{kh86}.

The columns toward stars in the central and southeastern parts of the
SMC (AV235, AV423, and Sk~188, stars 11,14, and 15) are lower than
those in the other parts of the galaxy. These regions have much less
star formation occurring \citep{k95}. While the columns are lower,
they are still as high as the Milky Way component along the same sight
lines. This is evidence for an extended component of
\ion{O}{6} in the SMC, which is locally enhanced by processes related
to star formation, such as SNRs. The \ion{O}{6} column density of hot gas in the
SMC rivals that of the Milky Way halo even in the absence of local
enhancement, which is perhaps surprising in a small galaxy with low
metallicity.

In Figure 4 the stars in and around NGC~346 stand out from the rest of
the points. If these points are excluded, the mean column density in
the SMC is log $N$(\ion{O}{6})$=14.45\pm^{0.10}_{0.14}$. This is about
1.4 times higher than the mean value of the column density
perpendicular to the Milky Way plane of log
$N_{\perp}$(\ion{O}{6})$=14.29$ \citep{s00}.  Not including the
contribution from the SNR toward AV229, the column densities toward
the NGC~346 stars, AV229, and AV232 are all very similar, almost
identical within the uncertainties of the measurements (counting the
NGC~346 stars as a single measurement).  In the rest of the SMC there
is more variation in the \ion{O}{6} column, in one case a factor of 2
difference between stars only a few arcminutes apart (AV83 and AV95,
stars 6 and 7), which is about the same as the distance between the
stars near NGC~346. The standard deviation in the column densities is
about 27\% (excluding the NGC~346 stars, AV229, and AV232), which is
somewhat lower than the deviation seen in the Milky Way \citep{s00}.
In general, the \ion{O}{6} column density appears to vary on both
large and small scales, with the exception being the smooth
distribution near NGC~346.

\section{Kinematics}
%section 4

We calculated the first and second moments of the \ion{O}{6}
1031.93~\AA\ absorption line profiles in the Milky Way and the SMC. The first moment
gives the average velocity of the line (weighted by the data values)
and the second moment gives the line width, equivalent to the standard
deviation ($\sigma$) if the line is Gaussian in shape. The line widths
given in Tables 2 and 3 were converted to FWHM values by multiplying
by 2.35. An initial concern was that measurement of these quantities
would be hindered by the blending of the SMC and Milky Way
components. In most cases the blue edge of the SMC absorption and the
red edge of the Milky Way absorption were chosen as the midpoint of
the shoulder between the two absorption peaks. Since part of the
low-velocity end of the line profile is missing due to blending, the
measured positions of the SMC lines may be weighted toward higher
velocities. However, comparison of the position measured in this
manner with the visually determined absorption minimum revealed that such
an effect did not occur, or was too small to be seen.

The \ion{O}{6} lines in the SMC are broader than the Milky Way
\ion{O}{6} lines. The FWHM ranges from $82-115$~\kms\ with a mean of
$94\pm9$~\kms, compared to the mean $59\pm11$~\kms\ for the Milky Way line
widths (which range from 40 to 66~\kms). Neither the NGC~346 stars nor
AV69 were included in the mean values. The SMC mean line width
corresponds to a temperature $T=3.1\times10^6$~K if the width is due
to thermal broadening alone, while the Milky Way mean line width
corresponds to $T=1.6\times10^6$~K. In collisional ionization
equilibrium, \ion{O}{6} peaks in abundance at $T=3\times10^5$~K
\citep{sd93}. If most of the gas is at this temperature, then either
large non-thermal motions are required to contribute to the line
broadening, or there are multiple components of \ion{O}{6} absorption
in both galaxies.

Figure 5 compares the \ion{O}{6} line profiles to those of
\ion{Fe}{2} $\lambda1144.94$.  In many of the sight lines there are two
low ionization absorption complexes in the SMC, a feature also seen
in $IUE$ spectra \citep{f85} and \ion{H}{1} 21cm studies
\citep{mn81,mfv88}. Most of the absorption from the low ionization gas
occurs in the lower velocity ($\sim+130$~\kms) component
\citep{dhfbs02}. The higher velocity ($\sim+180$~\kms) component is most
prominent in the southwest part of the Bar, and is not visible in the
spectra of stars near NGC~346. The spectrum of AV423 does not show the
double peaked profile, but the \ion{Fe}{2} absorption is as wide as
that in the AV235 spectrum, where the double peaked profile is
strong. Thus, it is possible that both components are present but
saturated toward AV423. This is not the case toward Sk~188, although
the \ion{Fe}{2} line is asymmetric toward the red side. Sk~188 is on
the far eastern part of the SMC, quite isolated from the other stars
in the sample.

The \ion{O}{6} absorption appears to lie close to the velocity of the
$+180$~\kms\ component of low-ionization absorption in those sight
lines where this component is visible, although there are a few cases
where the \ion{O}{6} absorption peaks between the two components
(AV26, AV14, AV15), and there are cases near NGC~346 where the peak is
redward of both components (AV321, AV378). In several cases (AV378,
AV321, AV83, and AV95) there may be structure in the \ion{O}{6}
absorption corresponding to the two low ionization components,
although the bulk of the absorption is still redward of the main low
ionization component at $+130$\kms. In the sight lines where the
$+180$\kms\ low ionization component is not visible, the
\ion{O}{6} is still systematically shifted to higher velocities than
the main component of low ionization absorption. This indicates that the
\ion{O}{6} absorbing gas has physically distinct properties from the gas
producing most of the low ionization absorption.

\section{Discussion}
%section 5

The observations described in this paper form a picture in which the
SMC has a substantial, widespread component of hot gas. On average the
\ion{O}{6} column density in the SMC is $\sim1.7$ times that of the
average column density perpendicular to the plane of the Milky Way of
log $N_{\perp}$(\ion{O}{6})$\approx14.29$ \citep{s00}. The
distribution of hot gas in the SMC appears to be strongly influenced
locally by nearby star formation. The largest \ion{O}{6} column
densities are measured along sight lines in or near star-forming
regions. The distribution of hot gas is strongly affected by star
formation in NGC~346, in particular, where the column density is
$\sim1.7$ times higher than the average for the rest of the SMC. The
\ion{O}{6} column density in the SMC {\it excluding} NGC~346 is
$\sim1.4$ times larger than the Galactic average.

Comparisons between the SMC and the Milky Way can be misleading
because the two galaxies have vastly different structures. It is not
clear that the SMC can be described as a planar system, in which an
inclination correction would be necessary. The system of carbon stars
in the SMC has been modeled as a plane with an inclination of
73$^\circ$ \citep{kdi00}. If this inclination holds for the hot gas
component of the SMC, then log $N_\perp$(\ion{O}{6})$=14.00$, and if
NGC~346 is excluded log $N_\perp$(\ion{O}{6})$=13.92$. The SMC
exhibits evidence for a large depth, with estimates ranging from 3.3
to 14 kpc (see {\it e.g.}, Welch et al. 1987; Groenewegen 2000), with
some estimates as high as 32 kpc \citep{mfv86}. The depth, optical
appearance, and the double-peaked \ion{H}{1} velocity profiles suggest
that a simple planar model is not applicable.  If the structure of the
SMC is truly irregular, it is probably more appropriate to use the
total column density, with no correction for inclination, for
comparison with the Milky Way.

Multiple supernovae and stellar winds associated with \ion{H}{2}
regions are a likely mechanism to produce the enhanced \ion{O}{6} seen
near star-forming regions in the SMC.  Supernovae are thought to
produce much of the hot gas in galaxies \citep{mo77,s90}, and one
supernova remnant observed by \fuse\ contains a large amount of
\ion{O}{6} \citep{hshb01}. In the Milky Way, \ion{O}{6} is higher than
the Galactic average on the sight line to 3C~273, which probes the
supernova remnant Loop IV \citep{s01}, although it is not enhanced
above the Scutum supershell \citep{ssrfs02}. Several giant \ion{H}{2}
regions contain X-ray emitting gas, so they probably contain coronal
phase gas as well \citep{wc95,w99}. It is important to note that the
average column density for the SMC may appear high compared to the
Milky Way because several of the SMC sight lines pass through
star-forming regions, while the Milky Way sight lines to extragalactic
sources typically do not.

One might expect the low metallicity of the SMC to result in a lower
\ion{O}{6} abundance. However, \cite{ec86} found that the metallicity
has little effect on the column density of hot, radiatively cooling
gas because the lower abundance is compensated by a longer cooling
time. If the \ion{O}{6} in the SMC arises in radiatively cooling hot
gas, our results confirm that low oxygen abundance does not reduce the
\ion{O}{6} abundance. In fact, it is conceivable that the lower
cooling efficiency could result in more hot gas, and thus a higher
observed \ion{O}{6} column density. This was suggested by \cite{w91}
to explain the soft X-ray halo.

The $+130$~\kms\ component of low-ionization gas is most often
interpreted as being in front of the $+180$~\kms\ component (see
Danforth et al. 2002). Extending this interpretation to the \ion{O}{6}
absorbing gas would mean that it is behind most of the cooler
gas. This interpretation is unsatisfactory for several reasons. The
\ion{O}{6} is seen at higher velocities than the bulk of the low
ionization material in nearly every sight line, requiring that every
star in the sample be associated with the weaker of the low ionization
components. In NGC~346, where the $+180$~\kms\ low ionization
component is not visible, the \ion{O}{6} is still redshifted compared
to the low ionization gas. There is no correlation between the
strength of the $+180$~\kms\ component and the \ion{O}{6} column
density; for example, the high velocity component is much more
pronounced in the AV83 spectrum than in AV15, yet the \ion{O}{6}
column toward AV15 is $\sim1.6$ times higher than toward AV83. A
correlation might be expected if the low ionization and high
ionization gas are cospatial. In many of the spectra the \ion{O}{6}
does not fall at the same velocity as the $+180$~\kms\ component.
Although these facts do not preclude a physical association between
the \ion{O}{6} and the $+180$~\kms\ low-ionization component, it is
also possible that the similarity of their velocity distributions is
coincidental.

\cite{ss97} suggest that the appearance of two \ion{H}{1} components in
the SMC is due to the presence of shells and supershells along the
line of sight, not due to distinct cloud complexes making up the
SMC. In this scenario the hot gas seen in \ion{O}{6} absorption could
lie between the two apparent components in velocity space if it fills
the interior of the shells. If these shells are most numerous along
the Bar where the star formation is concentrated, it would naturally
explain the higher \ion{O}{6} column densities measured there. To test
this idea we examined the sight lines that fall within shells listed
in the catalog of \cite{ss97}. These sight lines are: AV75, AV83, and
AV95 in the southwest (stars 8, 9, and 10), and AV321 and AV378 in the
north (stars 12 and 13). There are also shells in the direction of
AV14, AV26, and AV69, but we do not discuss these due to their uncertain
\ion{O}{6} profiles. All of these sight lines are in the Bar, but
there are other sight lines in the Bar and elsewhere in the SMC that
do not pass through any catalogued shell. \cite{ss97} point out that
shells may have been missed in regions with complex \ion{H}{1}
distribution. For the sight lines in the direction of a shell, the
shell velocities usually correspond to the peak of the \ion{O}{6}
absorption, although not in every case (see Figure 5). However, on all
five sight lines there is significant \ion{O}{6} absorption outside of
the velocity range of the expanding shells. Furthermore, many of the
sight lines are not in the direction of any shell. We conclude that
hot gas in expanding shells produces some of the \ion{O}{6}
absorption, but the bulk is produced in gas outside of the shells.

A galactic fountain might also explain the observations. In this
scenario correlated supernovae produce a flow of very hot gas into the
halo, which then cools and falls back on the main body of the
galaxy. Absorption from \ion{O}{6} is produced as the gas cools
through temperatures of $(1-3)\times10^5$~K. If at this point the gas
is falling back toward the SMC, a shift toward higher velocities from
the low-ionization gas would be expected, as is seen in the data. This
scenario might also explain the correlation with star formation, if
the expelled gas remains close to the site of star formation. This
might not be expected in a rotating disk system like the Milky Way,
but would be more likely in the SMC, which appears to lack a coherent
direction of rotation. \cite{w91} found diffuse X-ray emission in the
SMC with spectral properties that favor an origin in hot gas, leading
them to suggest a fountain model for the SMC. The
\ion{O}{6} observations are consistent with this
interpretation. This model is closely related to the supershell
scenario discussed above, with the key difference being whether the
hot gas is located in the main body of the SMC or in the
halo. \cite{in87} and \cite{ni89} state that a burst of star formation
in a dwarf galaxy will form a superbubble that will easily break out
into the halo because the gravitational potential is small, so it is
reasonable to expect a fountain to exist in a star-forming dwarf
galaxy like the SMC.

Using the model of \cite{ec86} for gas radiatively cooling from
T~$\ga$~10$^6$~K, it is possible to roughly
estimate the mass flow rate implied by the \ion{O}{6} column
density. The expression for the
mass flow rate {\it \.M}/$\Omega$ is
\begin{equation}
\frac{\dot M}{\Omega} \approx(\mu m_{\mathrm H})~n_{\mathrm H0} \left( \frac{\dot N}{n_{\mathrm H0}} \right) ,
\end{equation}
where $\mu m_{\mathrm H}$ is the mean mass per atom, $n_{\mathrm H0}$
is the initial density of ionized hydrogen in the cooling gas,
$\Omega$ is the surface area of the fountain region, and $\dot N$ is
the flux if cooling gas in units of hydrogen ions
cm$^{2}s^{-1}$. Edgar \& Chevalier find that under isobaric conditions
$\dot N/n_{\mathrm H0} \sim 2.5\times10^6
[N_{\perp}$(\ion{O}{6})$/10^{14}\mathrm{cm}^{-2}]$. Using the mean \ion{O}{6}
column density in Table 4 leads to a mass flow per unit surface area
of
\begin{equation}
\frac{\dot M}{\Omega}  \approx 2\times10^{-2} \left( \frac{n_{\mathrm H0}}{10^{-2}~ \mathrm{cm}^{-3}} \right) \mathrm{M}_{\odot}~ \mathrm{yr}^{-1}~ \mathrm{kpc}^{-2}.
\end{equation}
We leave $n_{\mathrm H0}$ as an unknown, as there is no reliable way
to estimate this for the SMC. Similarly, an estimate of the surface
area of the SMC is very difficult to make in light of the irregular
morphology and unknown orientation of the galaxy. Our estimate for the
mass flow per unit area is probably similar to the number that would
be derived for the Milky Way, since the \ion{O}{6} column densities
are similar (to within a factor of 2). A more detailed discussion of
this type of calculation can be found in \cite{hsfs02}.

If the \ion{O}{6} is in front of both low ionization components, it is
falling toward the SMC. This might occur if the hot gas were stripped
out of the SMC by the gravitational interaction with the Milky Way,
and is now falling back on the SMC. \cite{sem00} found that the
Magellanic Stream contains a high column of \ion{O}{6} (log
$N$(\ion{O}{6})$\approx14.62$), comparable to the columns seen in the
SMC. This would naturally explain the observed velocity shifts. It
does not, however, explain the observed spatial correlation between
high \ion{O}{6} column density and star-forming regions. 

A clearer understanding of the origin of hot gas in the SMC might be
gained by comparing \ion{O}{6} to other high ions, such as \ion{C}{4},
\ion{Si}{4}, and \ion{N}{5}. Unfortunately, very little work has been
done on these ions in the SMC with $HST$, so most of the available
information is from $IUE$ spectra. The low signal-to-noise of the $IUE$
spectra make the determination of accurate column densities
difficult. Table 5 lists the ratios of the column densities of high ions
for the four \fuse\ stars with reliable $IUE$ measurements. Of the
four sight lines, the gas toward AV235 appears to be in the highest
ionization state. However, AV229 and AV232 are in a star-forming
region where appreciable \ion{C}{4} and \ion{Si}{4} may be produced
locally. Sk~188 is a Wolf-Rayet star, which might also lead to
\ion{C}{4} and \ion{Si}{4} production. Thus, the limited $IUE$ data for
\ion{C}{4} and \ion{Si}{4} do not allow a discrimination
between the possible origins of the highly ionized gas in the SMC.

A STIS spectrum of one SMC sight line, AV229 (HD 5980), has been
analyzed by \cite{k01} and \cite{hshb01}. This sight line passes
through an SNR in the SMC, but the SNR absorption is well separated
from the SMC absorption. \cite{hshb01} showed that the
\ion{C}{4}/\ion{O}{6} and \ion{C}{4}/\ion{N}{5} ratios in the general
ISM along this sight line are in the range of the Milky Way halo
ratios, implying that the processes producing the hot gas in the two
galaxies are similar. The ratios in the SNR absorption were found to
agree well with those predicted for evolved supernova remnants, and
were different from the general ISM ratios in the SMC, suggesting that
not all of the hot gas is in the form of cooling SNRs. AV229 is a
Luminous Blue Variable/Wolf-Rayet binary star and is in a large
\ion{H}{2} region, so there may be local factors affecting the
\ion{C}{4} and \ion{O}{6} abundance. The \fuse\ data toward Sk~108
indicate that the \ion{C}{4}/\ion{O}{6} ratio is several times higher
than the Milky Way halo, suggesting that the ratio varies widely in
the SMC \citep{m01}. A systematic survey of high ion absorption in the SMC with
STIS would be useful for studying the origin of highly ionized
gas.

\section{Conclusions}
%section 6

In this paper we have described the properties of \ion{O}{6}
absorption along 18 sight lines in the Small Magellanic Cloud, using
far ultraviolet spectra obtained with \fuse. The main results are:

\begin{enumerate}

\item{Absorption from \ion{O}{6} is seen along every sight line in the
SMC. The sight lines range in environment from star-forming regions
(\eg, NGC~346) to the field (\eg, AV423), and includes the periphery
of the SMC (Sk~188). The occurrence of \ion{O}{6} absorption in every
direction indicates the presence of a widespread component of coronal
temperature gas in the SMC. }

\item{The average column density is log $N$(\ion{O}{6})
$\sim14.53$. This is about 1.7 times higher then the Milky Way average
of the column density perpendicular to the plane of log
$N$(\ion{O}{6}) $\sim14.29$ \citep{s00}. However, if NGC~346 is
excluded, the average for the rest of the SMC falls to $\sim14.45$,
only 1.4 times higher than the Galactic average. }

\item{The column density is correlated with position in the SMC. The
highest values are seen toward the star-forming region NGC~346, and
to a lesser extent toward the star-forming regions in the southwestern
end of the Bar. Lower values are seen in the central region and in the
Wing, where there is little star formation activity. This suggests
that local processes strongly affect the hot gas distribution in the
SMC, particularly in NGC~346.}

\item{The \ion{O}{6} line widths are broader than expected for pure
thermal broadening, indicating that either non-thermal motions are
prevalent, or that there are multiple components of hot gas at
different velocities. The SMC \ion{O}{6} line widths are broader than
the Milky Way \ion{O}{6} lines.}

\item{The kinematics suggest that the \ion{O}{6} is not associated with
the gas producing the bulk of the low ionization absorption at
$\sim+130$~\kms. The \ion{O}{6} velocity is closer that of the
$\sim+180$~\kms\ low ionization component, but evidence for a physical
association is not conclusive. We discuss several mechanisms that may
explain the observations, with superbubbles and/or a galactic
fountain being the most probable.}

\item{If a galactic fountain is responsible for producing the hot gas,
the rate of mass flow per unit surface area of the flow region is
$\dot M/\Omega \sim 2\times10^{-2}$ M$_{\odot}$ yr$^{-1}$ kpc$^{-2}$,
assuming an initial gas density of 10$^{-2}$ cm$^{-2}$, which is very
uncertain. Since the \ion{O}{6} column density in the SMC is within a
factor of 2 of that of the Milky Way, the mass flow rates are probably
similar if other conditions are the same.}

\end{enumerate}

\acknowledgments

We thank the anonymous referee for helpful comments. This work is
based on data obtained for the Guaranteed Time Team by the
NASA-CNES-CSA FUSE mission operated by the Johns Hopkins
University. Financial support has been provided by NASA contract
NAS5-32985. The SMC H$\alpha$ image was taken from the Southern
H-Alpha Sky Survey Atlas (SHASSA), which is supported by the National
Science Foundation.  KRS and JCH acknowledge partial financial support
from Long Term Space Astrophysics grant NAG5-3485.

%% Figures

\begin{figure}
\epsscale{0.80}
\plotone{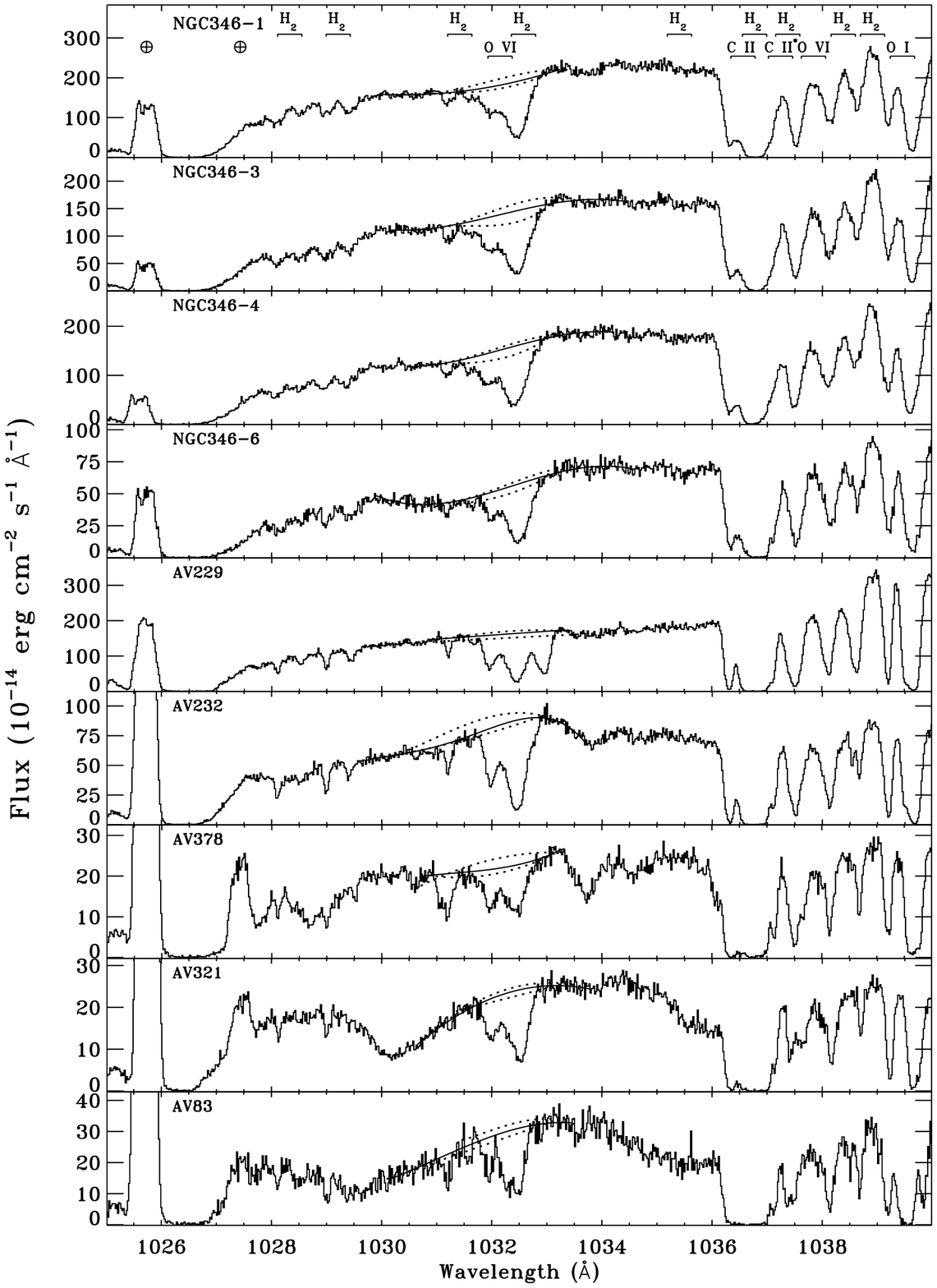}
\caption{The observed \FUSE\ spectra of the sample stars near the
O~VI lines. Atomic and H$_2$ lines in the Galactic and SMC rest
frames are identified in the top panel. The adopted continuum over
the O~VI 1031.93~\AA\ line is shown as a solid line, and the maximum and
minimum continua used to determine the continuum placement uncertainty
are shown as dashed lines. The spectra of AV14, AV26, and AV69 are separated
from the rest as they may suffer from contamination by stellar
absorption. }
\end{figure}

\begin{figure}
\epsscale{0.80}
\plotone{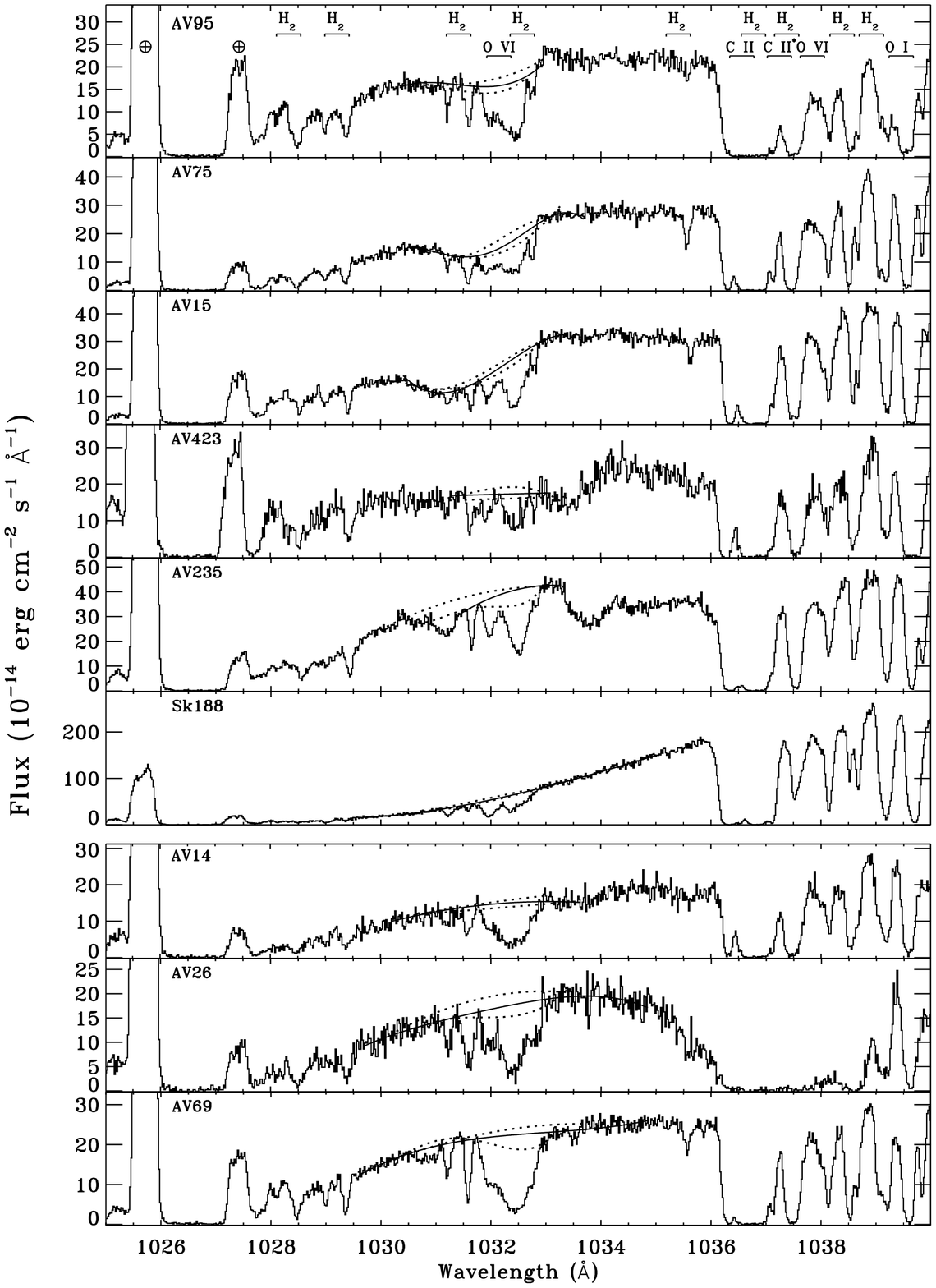}
\end{figure}

\begin{figure}
\epsscale{0.80}
\plotone{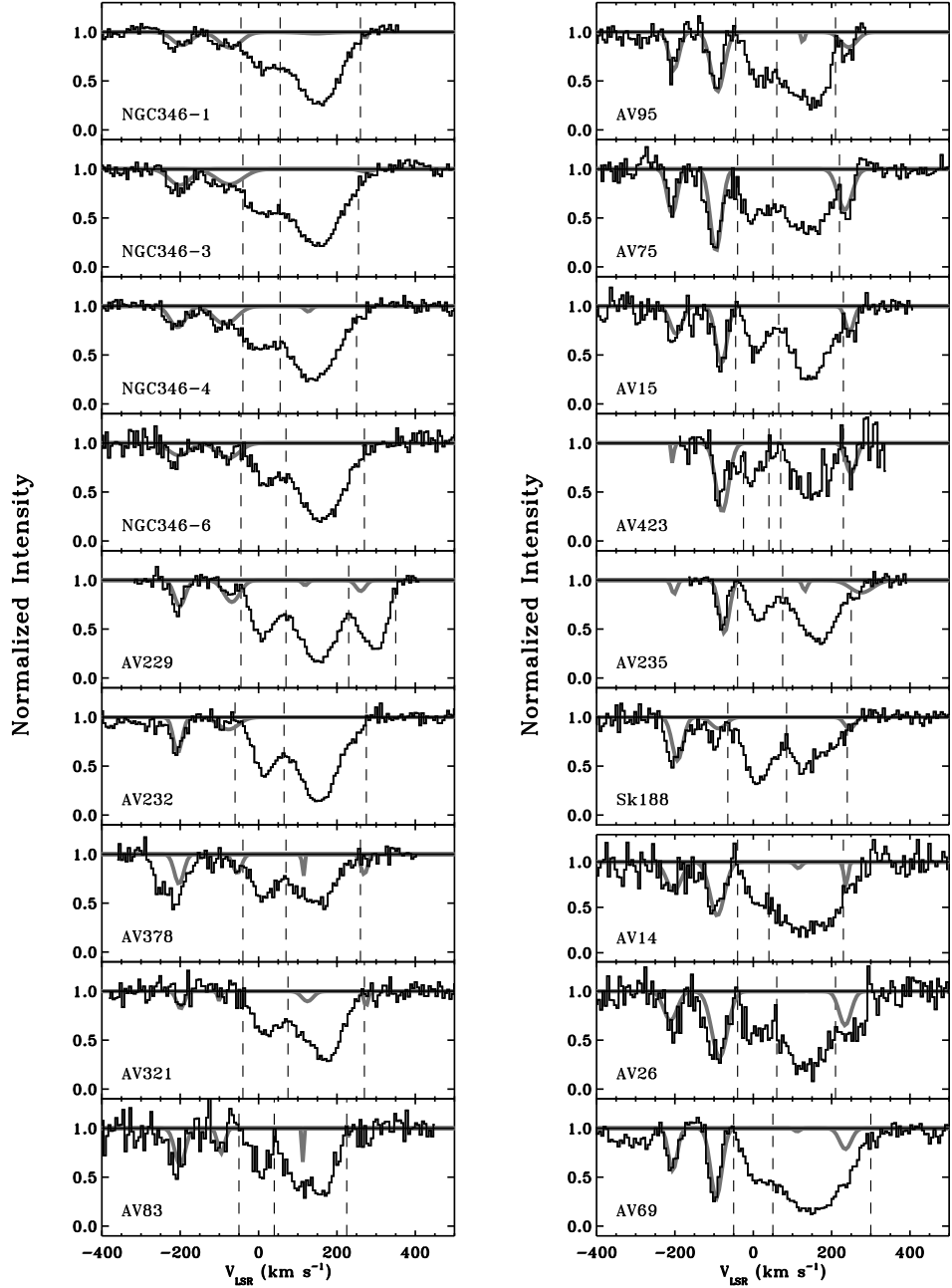}
\caption{The continuum-normalized O VI profiles. The thick grey lines are
the predicted strengths of the H$_2$ 6-0~P(3) 1031.19~\AA\ and
6-0~R(4) 1032.35~\AA\ lines. The predictions are based on nearby
lines from the same rotational level. The vertical dashed lines show
the velocity integration limits used to measure equivalent widths and
column densities. The profiles for AV14, AV26, and AV69 are separated
from the rest as they may suffer from contamination by stellar
absorption.}
\end{figure}

\begin{figure}
\epsscale{1.0}
\plotone{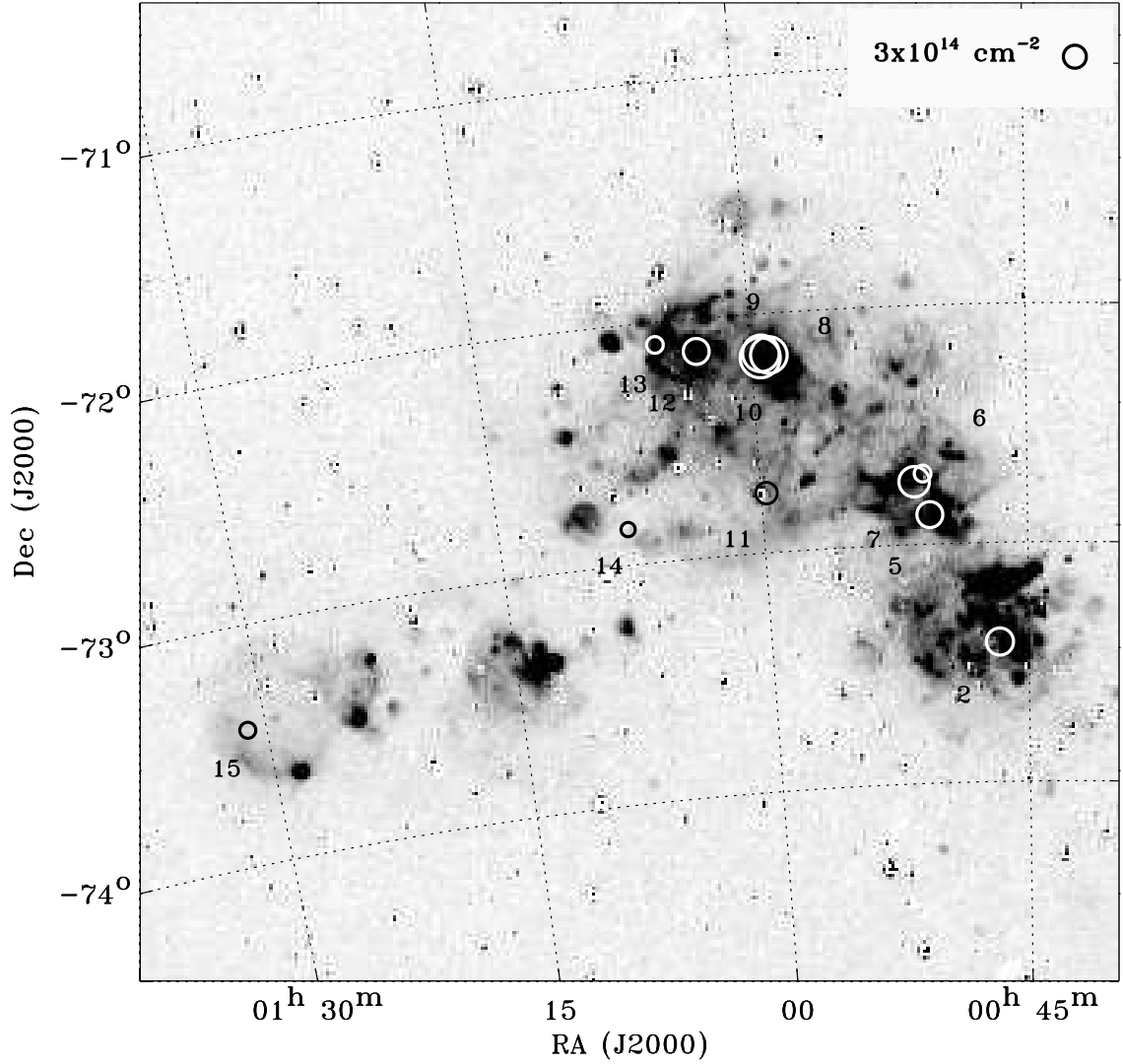}
\caption{H$\alpha$ image of the SMC (Gaustad et al. 2001) with the
positions of the sight lines marked. The radius of the circle used to
mark each star is linearly proportional to the column density of O VI
in the SMC. A circle corresponding to $N$(O VI)=$3\times10^{14}$ cm$^{-2}$ is
shown in the upper right corner. The numbers given beside each circle
correspond to the identifications listed in Table 1. AV14, AV26, and
AV69 are not shown as the measured column densities toward these stars
are unreliable. They are located in the southwestern end of the Bar,
near star 5 (AV75). The four stars in NGC~346 are labeled as one (star
8) as they are very close together.}
\end{figure}

\begin{figure}
\epsscale{0.85}
\plotone{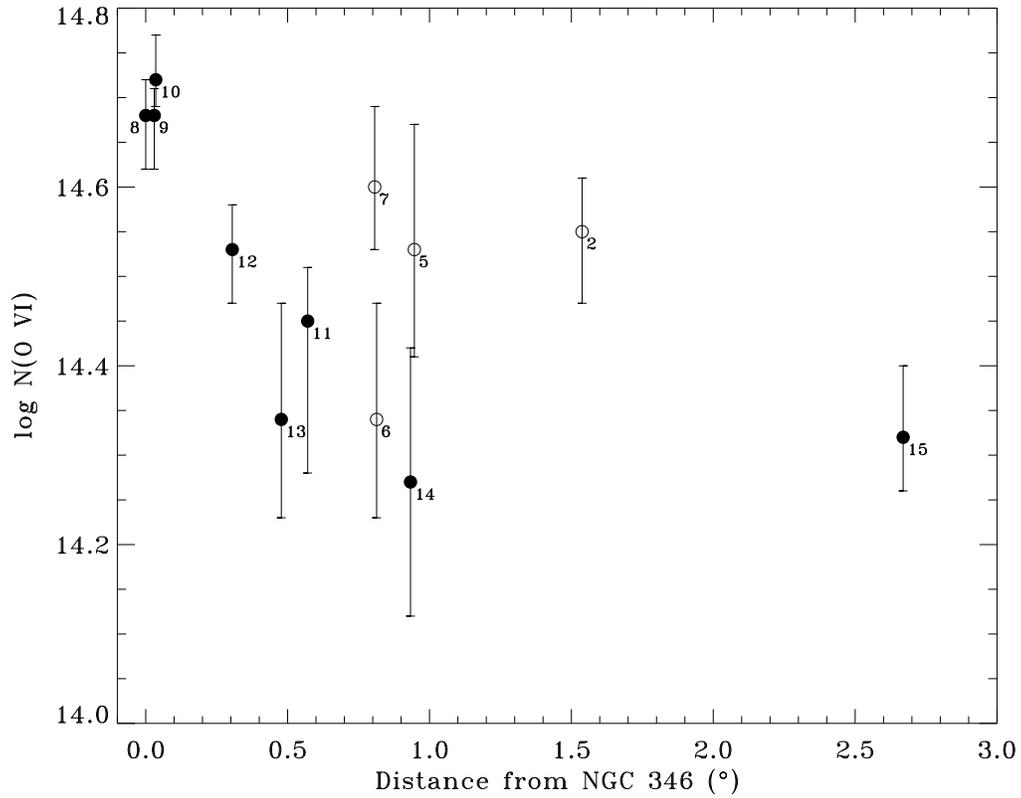}
\caption{The measured O VI column densities plotted as a function of
distance from NGC~346. The average value for the stars in NGC~346 is
shown (point 8). The column density toward AV229 (point 9) does not
include the contribution from the SNR observed along the sight
line. The open circles represent the stars in the southwestern end
of the Bar, where star formation is occurring. The error bars shown
include both random and systematic (continuum fit)
uncertainty. Unreliable values for several stars (AV69, AV14, and
AV26) are not shown.}
\end{figure}

\begin{figure}
\epsscale{0.7}
\plotone{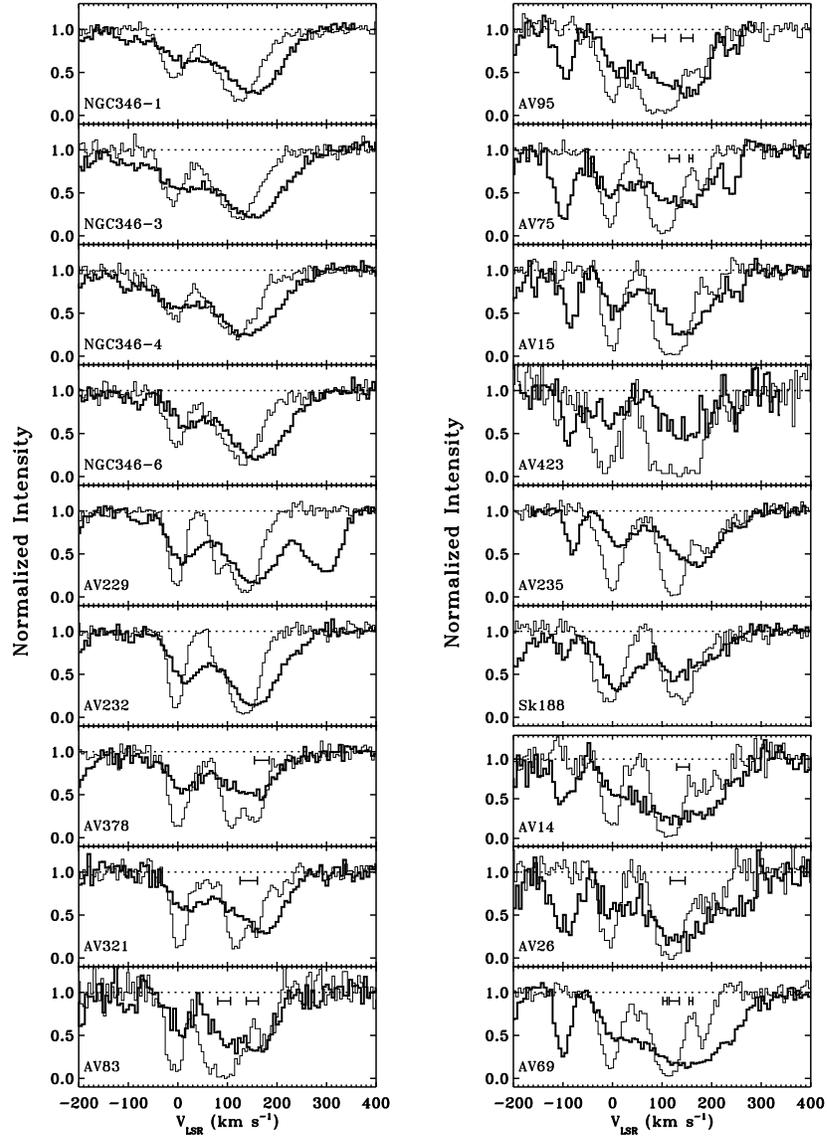}
\caption{The continuum-normalized O~VI profiles compared to
Fe~II 1144.94~\AA\ absorption in velocity space. The O~VI absorption
is shown by the thick line, while the Fe~II profile is shown by the
thin line. The velocity scale is set so that the Milky Way Fe~II
absorption falls at 0\kms. The SMC absorption lies between $\sim+100$
and $+200$~\kms, and often consists of two components in the Fe~II
line. The absorption at $-100$ and $+260$ \kms\ seen in some of the O
VI spectra is caused by the 6-0~P(3) and 6-0~R(4) levels of H$_2$
in the SMC, respectively. The O VI absorption at $+300$~\kms\ toward
AV229 is caused by a supernova remnant in the line of sight (Hoopes et
al. 2001). The horizontal error bars denote the velocities of
identified shells along the sight line, from the catalog of
Staveley-Smith et al. (1997). The left and right ends of each error
bar mark the velocities of the approaching and receding sides of
each shell, respectively. The profiles for AV14, AV26, and AV69 are
separated from the rest as they may suffer from contamination by
stellar absorption.}
\end{figure}

%% Tables

\clearpage

\begin{deluxetable}{lccccccccccc}
\tabletypesize{\scriptsize}
\tablewidth{0pc}
\tablenum{1}
\tablecaption{Information on Observed Stars} 
\tablehead{
\colhead{ID\tablenotemark{a}}&\colhead{Star}& \colhead{Alternate}& \colhead{$l$}& \colhead{$b$}
& \colhead{$\alpha_{2000}$}& \colhead{$\delta_{2000}$}& \colhead{Type\tablenotemark{b}} 
& \colhead{HII Region\tablenotemark{c}}\\
\colhead{}&\colhead{}& \colhead{Names}& \colhead{($^{\circ}$)}& \colhead{($^{\circ}$)}
& \colhead{(h~~m~~s)}& \colhead{($^{\circ}~~^{\prime}~~^{\prime\prime}$)}& \colhead{} & \colhead{}
}
\startdata
1  & AV14    & Sk~9           & 303.43 & $-$44.02 & 00:46:32.66 & $-$73:06:05.6 & O3-4V          & N12     \\
2  & AV15    & Sk~10          & 303.40 & $-$43.71 & 00:46:42.19 & $-$73:24:54.7 & O6.5II(f)      & \nodata \\
3  & AV26    & Sk~18          & 303.30 & $-$43.99 & 00:47:50.07 & $-$73:08:20.7 & O7III          & N19     \\
4  & AV69    & Sk~34          & 303.05 & $-$44.24 & 00:50:17.40 & $-$72:53:29.9 & OC7.5III((f))  & N36     \\
5  & AV75    & Sk~38          & 303.02 & $-$44.25 & 00:50:32.50 & $-$72:52:36.2 & O5III(f+)      & N36     \\
6  & AV83    & \nodata       & 302.99 & $-$44.42 & 00:50:52.01 & $-$72:42:14.5 & O7Iaf+         & \nodata \\
7  & AV95    & \nodata       & 302.94 & $-$44.39 & 00:51:21.54 & $-$72:44:12.9 & O7III((f))     & \nodata \\
8  & NGC~346-6  & MPG\#324      & 302.12 & $-$44.94 & 00:58:57.74 & $-$72:10:33.6 & O4V((f))       & N66     \\
8  & NGC~346-4  & MPG\#342      & 302.11 & $-$44.94 & 00:59:00.39 & $-$72:10:37.9 & O5-6V          & N66     \\
8  & NGC~346-3  & MPG\#355      & 302.11 & $-$44.94 & 00:59:01.09 & $-$72:10:28.2 & O2III(f$^{\ast}$)& N66     \\
8  & NGC~346-1  & MPG\#435      & 302.11 & $-$44.94 & 00:59:04.81 & $-$72:10:24.8 & O4III(n)(f)    & N66     \\
9  & AV229   & Sk~78, HD5980  & 302.07 & $-$44.95 & 00:59:26.70 & $-$72:09:55.0 & OB?+WN3        & N66B    \\
10 & AV232   & Sk~80          & 302.06 & $-$44.94 & 00:59:30.00 & $-$72:11:00.0 & O7Iaf+         & N66     \\
11 & AV235   & Sk~82          & 302.07 & $-$44.37 & 00:59:42.00 & $-$72:45:00.0 & B0Iw           & \nodata \\
12 & AV321   & \nodata       & 301.68 & $-$44.96 & 01:02:57.04 & $-$72:08:09.3 & B0IIIw         & \nodata \\
13 & AV378   & \nodata       & 301.44 & $-$45.00 & 01:05:09.44 & $-$72:05:35.0 & O8V            & \nodata \\
14 & AV423   & Sk~132         & 301.26 & $-$44.23 & 01:07:40.43 & $-$72:50:59.6 & O9.5II(n)      & \nodata \\
15 & Sk~188   & AB8           & 299.06 & $-$43.41 & 01:31:06.00 & $-$73:26:00.0 & WO4+O7III      & \nodata \\
\enddata
\tablenotetext{a}{Star identification number used in Figures 3 and 4.}
\tablenotetext{b}{Spectral types were taken from Walborn 1977; Barlow \& Hummer 1982; Walborn \& Blades 1986; Garmany et al. 1987; Walborn et al. 1995, 2000, 2001a, 2001b.}
\tablenotetext{c}{The HII region designations are from Henize 1956.}
\end{deluxetable}

\begin{deluxetable}{llcc}
\tablewidth{0pc}
\tabletypesize{\scriptsize}
\tablenum{2}
\tablecaption{FUSE Observations} 
\tablehead{\colhead{Star} & \colhead{Date} & \colhead{FUSE dataset ID\tablenotemark{a}}& \colhead{Exposure Time}\\
\colhead{} & \colhead{} & \colhead{}& \colhead{(ks)} }
\startdata
AV14    & 2000 Jul 1  & P11753 & 6.8  \\
AV15    & 2000 May 30 & P11501 & 14.6 \\
AV26    & 2000 Jul 2  & P11760 & 4.0  \\
AV69    & 2000 May 31 & P11503 & 17.6 \\
AV75    & 2000 May 31 & P11504 & 14.4 \\
AV83    & 2000 Jul 2  & P11762 & 4.0  \\
AV95    & 2000 May 31 & P11505 & 14.3 \\
NGC~346-6  & 2000 Oct 1  & P11756 & 5.9  \\
NGC~346-4  & 2000 Sep 30 & P11757 & 5.0  \\
NGC~346-3  & 2000 Oct 1  & P11752 & 5.1  \\
NGC~346-1  & 2000 Oct 11 & P11755 & 5.6  \\
AV229   & 2000 Jul 2  & P10301 & 5.5  \\
AV232   & 2000 Jul 2  & P10302 & 11.7 \\
AV235   & 2000 Jul 2  & P10303 & 16.2 \\
AV321   & 2000 Jun 1  & P11506 & 16.9 \\
AV378   & 2000 Jun 1  & P11507 & 14.7 \\
AV423   & 2000 Oct 10 & P11767 & 3.5  \\
Sk~188   & 2000 Oct 11 & P10306 & 6.9  \\
\enddata
\tablenotetext{a}{The data are archived in the Multimission Archive at the Space Telescope Science Institute.}
\end{deluxetable}

\begin{deluxetable}{lcccccc}
\tabletypesize{\scriptsize}
\tablecolumns{7}
\tablewidth{0pc}
\tablenum{3}
\tablecaption{Measured Properties of the Galactic O VI Absorption} 
\tablehead{
\colhead{ID\tablenotemark{a}} &\colhead{Star} &\colhead{Velocity range} & \colhead{$W_\lambda$(O VI)\tablenotemark{b}} &
\colhead{log $N$(O VI)\tablenotemark{b}} & \colhead{$<v>$\tablenotemark{c}} & \colhead{$\Delta v$\tablenotemark{c}} \\
\colhead{}     &\colhead{}     & \colhead{(\kms)} &\colhead{(m\AA)}          &\colhead{(cm$^{-2}$)} & \colhead{(\kms)} & 
\colhead{(\kms)}
}
\startdata
1  & AV14\tablenotemark{d} & $-40$, $+40$ & $102\pm12\pm^{4}_{19}$  & $14.02\pm0.08\pm^{0.03}_{0.08}$ & $11\pm3$ & $47\pm5$ \\
2  & AV15                  & $-45$, $+65$ & $110\pm12\pm^{26}_{20}$ & $14.04\pm0.06\pm^{0.11}_{0.10}$ & $16\pm4$ & $54\pm7$ \\
3  & AV26\tablenotemark{d} & $-40$, $+60$ & $127\pm15\pm^{30}_{24}$ & $14.12\pm0.07\pm^{0.12}_{0.10}$ & $13\pm4$ & $59\pm7$ \\
4  & AV69\tablenotemark{d} & $-50$, $+50$ & $144\pm7\pm^{10}_{12}$  & $14.20\pm0.04\pm^{0.04}_{0.05}$ & $13\pm3$ & $56\pm7$ \\
5  & AV75                  & $-40$, $+50$ & $134\pm14\pm^{27}_{17}$ & $14.15\pm0.07\pm^{0.12}_{0.07}$ &  $8\pm2$ & $56\pm5$ \\
6  & AV83                  & $-50$, $+40$ &  $84\pm11\pm^{16}_{18}$ & $13.93\pm0.07\pm^{0.08}_{0.14}$ &  $4\pm4$ & $42\pm7$ \\
7  & AV95                  & $-45$, $+60$ & $130\pm20\pm^{18}_{25}$ & $14.14\pm0.07\pm^{0.07}_{0.11}$ & $19\pm5$ & $56\pm9$ \\
8  & NGC~346-1             & $-45$, $+55$ & $115\pm10\pm^{17}_{10}$ & $14.05\pm0.04\pm^{0.08}_{0.05}$ & $10\pm3$ & $64\pm9$ \\ 
8  & NGC~346-3             & $-40$, $+55$ & $140\pm6\pm^{12}_{26}$  & $14.16\pm0.04\pm^{0.05}_{0.11}$ &  $9\pm2$ & $61\pm7$ \\ 
8  & NGC~346-4             & $-40$, $+55$ & $137\pm5\pm^{6}_{24}$   & $14.15\pm0.04\pm^{0.02}_{0.11}$ &  $9\pm2$ & $64\pm9$ \\
8  & NGC~346-6             & $-40$, $+65$ & $112\pm8\pm^{6}_{27}$   & $14.04\pm0.05\pm^{0.03}_{0.13}$ & $22\pm3$ & $61\pm7$ \\
9  & AV229                 & $-45$, $+70$ & $164\pm5\pm^{10}_{39}$  & $14.25\pm0.03\pm^{0.03}_{0.06}$ & $17\pm2$ & $64\pm5$ \\
10 & AV232                 & $-60$, $+65$ & $155\pm6\pm^{31}_{12}$  & $14.23\pm0.03\pm^{0.09}_{0.05}$ & $19\pm3$ & $61\pm5$ \\
11 & AV235                 & $-40$, $+75$ & $101\pm12\pm^{22}_{30}$ & $13.98\pm0.06\pm^{0.12}_{0.16}$ & $22\pm4$ & $64\pm9$ \\
12 & AV321                 & $-40$, $+75$ & $130\pm8\pm^{10}_{11}$  & $14.11\pm0.05\pm^{0.04}_{0.04}$ & $24\pm3$ & $66\pm5$ \\
13 & AV378                 & $-40$, $+70$ & $124\pm12\pm^{24}_{16}$ & $14.09\pm0.06\pm^{0.10}_{0.08}$ & $17\pm3$ & $64\pm7$ \\
14 & AV423                 & $-25$, $+40$ &  $61\pm11\pm^{13}_{14}$ & $13.77\pm0.10\pm^{0.10}_{0.12}$ &  $4\pm3$ & $40\pm5$ \\
15 & Sk~188                 & $-65$, $+85$ & $215\pm9\pm^{23}_{12}$  & $14.39\pm0.03\pm^{0.05}_{0.04}$ & $17\pm2$ & $78\pm5$ \\
Mean\tablenotemark{e} & \nodata  & \nodata      & $126\pm37$         & $14.12\pm^{0.13}_{0.18}$          & $15\pm6$ & $58\pm10$\\
Mean\tablenotemark{f} & \nodata  & \nodata      & $128\pm39$         & $14.12\pm^{0.14}_{0.20}$          & $15\pm7$ & $58\pm10$\\
\enddata
\tablenotetext{a}{Star identification number used in Figures 3 and 4.}
\tablenotetext{b}{The first uncertainty includes statistical and
velocity range errors, and the second is due to continuum placement
errors.}
\tablenotetext{c}{The mean velocity is the first moment of the line, so
the statistics are weighted by the data values. The width is the
second moment multiplied by 2.35, or roughly the FWHM.}
\tablenotetext{d}{Stellar absorption may affect the measurement of the
interstellar absorption toward AV69, AV14, and AV26.}
\tablenotetext{e}{The NGC~346 values were averaged together and treated
as one measurement in the global mean. AV69 is not included in this
mean.}
\tablenotetext{f}{The NGC~346 values were averaged together and treated
as one measurement in the global mean. AV69, AV14, and AV26 are not
included in this mean.}
\end{deluxetable}

\begin{deluxetable}{lcccccc}
\tabletypesize{\scriptsize}
\tablecolumns{7}
\tablewidth{0pc}
\tablenum{4}
\tablecaption{Measured Properties of the SMC O VI Absorption} 
\tablehead{
\colhead{ID\tablenotemark{a}} &\colhead{Star} &\colhead{Velocity range} & \colhead{$W_\lambda$(O VI)\tablenotemark{b}}
&\colhead{log $N$(O VI )\tablenotemark{b}} & \colhead{$<v>$\tablenotemark{c}} 
& \colhead{$\Delta v$\tablenotemark{c}} \\
\colhead{}  &\colhead{}  & \colhead{(\kms)} &\colhead{(m\AA)} &\colhead{(cm$^{-2}$)} & \colhead{(\kms)} & \colhead{(\kms)}
}
\startdata
\cutinhead{ Northern Bar near NGC~346}
8  & NGC~346-1              & $+55$, $+260$ & $349\pm10\pm^{26}_{15}$ & $14.63\pm0.02\pm^{0.04}_{0.02}$ & $148\pm3$ & $106\pm5$ \\ 
8  & NGC~346-3              & $+55$, $+255$ & $395\pm8\pm^{24}_{43}$  & $14.71\pm0.02\pm^{0.04}_{0.07}$ & $148\pm2$ & $108\pm5$ \\ 
8  & NGC~346-4              & $+55$, $+250$ & $361\pm8\pm^{31}_{14}$  & $14.67\pm0.02\pm^{0.01}_{0.07}$ & $142\pm2$ & $104\pm5$ \\
8  & NGC~346-6              & $+65$, $+270$ & $377\pm10\pm^{15}_{30}$ & $14.69\pm0.02\pm^{0.02}_{0.05}$ & $160\pm2$ & $104\pm5$ \\
9  & AV229                  & $+70$, $+350$ & $555\pm12\pm^{12}_{34}$ & $14.86\pm0.03\pm^{0.02}_{0.05}$ & \nodata   & \nodata   \\
9  & AV229\tablenotemark{d} & $+70$, $+230$ & $345\pm7\pm^{9}_{17}$   & $14.68\pm0.02\pm^{0.01}_{0.04}$ & $151\pm2$ & $97\pm7$  \\
10 & AV232                  & $+65$, $+275$ & $384\pm8\pm^{18}_{10}$  & $14.72\pm0.02\pm^{0.03}_{0.01}$ & $153\pm2$ & $97\pm5$  \\
\cutinhead{ Northern Bar outside NGC~346 }
13 & AV378                  & $+70$, $+260$ & $211\pm17\pm^{31}_{17}$ & $14.34\pm0.05\pm^{0.08}_{0.06}$ & $149\pm4$ & $101\pm9$ \\
12 & AV321                  & $+75$, $+270$ & $290\pm9\pm^{13}_{18}$  & $14.53\pm0.03\pm^{0.02}_{0.03}$ & $160\pm2$ & $94\pm5$  \\
\cutinhead{ Southwestern Bar }
6  & AV83                   & $+40$, $+225$ & $296\pm15\pm^{20}_{20}$ & $14.34\pm0.05\pm^{0.08}_{0.06}$ & $133\pm3$ & $97\pm5$  \\
7  & AV95                   & $+60$, $+210$ & $313\pm18\pm^{22}_{15}$ & $14.60\pm0.04\pm^{0.05}_{0.03}$ & $135\pm3$ & $89\pm7$  \\
4  & AV69\tablenotemark{e}  & $+50$, $+300$ & $544\pm11\pm^{22}_{53}$ & $14.90\pm0.02\pm^{0.03}_{0.07}$ & $155\pm2$ & $131\pm5$ \\
5  & AV75                   & $+50$, $+220$ & $298\pm16\pm^{44}_{42}$ & $14.53\pm0.04\pm^{0.10}_{0.08}$ & $133\pm3$ & $97\pm5$  \\
3  & AV26\tablenotemark{e}  & $+60$, $+210$ & $344\pm15\pm^{22}_{27}$ & $14.68\pm0.05\pm^{0.05}_{0.05}$ & $135\pm3$ & $87\pm5$  \\
1  & AV14\tablenotemark{e}  & $+40$, $+230$ & $435\pm15\pm^{11}_{19}$ & $14.77\pm0.04\pm^{0.02}_{0.03}$ & $134\pm3$ & $115\pm5$ \\
2  & AV15                   & $+65$, $+230$ & $291\pm11\pm^{14}_{26}$ & $14.55\pm0.03\pm^{0.03}_{0.05}$ & $146\pm2$ & $85\pm5$  \\
\cutinhead{ Center }
14 & AV423                  & $+70$, $+230$ & $174\pm19\pm^{35}_{30}$ & $14.27\pm0.06\pm^{0.09}_{0.09}$ & $146\pm5$ & $82\pm7$  \\
11 & AV235                  & $+75$, $+250$ & $256\pm14\pm^{7}_{56}$  & $14.45\pm0.04\pm^{0.02}_{0.13}$ & $163\pm4$ & $94\pm9$  \\
\cutinhead{ Wing }
15 & Sk~188                  & $+85$, $+240$ & $202\pm7\pm^{17}_{9}$   & $14.32\pm0.03\pm^{0.05}_{0.03}$ & $153\pm2$ & $92\pm5$  \\
\cutinhead{ Global Properties }
Mean\tablenotemark{f}  & \nodata  & \nodata   & $300\pm73$         & $14.56\pm^{0.14}_{0.20}$        & $145\pm10$& $94\pm8$  \\
Mean\tablenotemark{g}  & \nodata  & \nodata   & $286\pm66$         & $14.53\pm^{0.13}_{0.19}$        & $147\pm10$& $93\pm6$  \\
\enddata
\tablenotetext{a}{Star identification number used in Figures 3 and 4.}
\tablenotetext{b}{The first uncertainty includes statistical and
velocity range errors, and the second is due to continuum placement
errors.}
\tablenotetext{c}{The mean velocity is the first moment of the line, so
the statistics are weighted by the data values. The width is the
second moment multiplied by 2.35, or roughly the FWHM.}
\tablenotetext{d}{This measurement excludes the SNR.}
\tablenotetext{e}{Stellar absorption may affect the measurement of the
interstellar absorption toward AV69, AV14, and AV26.}
\tablenotetext{f}{The NGC~346 values were averaged together and treated
as one measurement in the global mean. AV69 is not included in this
mean. The SNR is excluded.}
\tablenotetext{g}{The NGC~346 values were averaged together and treated
as one measurement in the global mean. AV69, AV14, and AV26 are not
included in this mean. The SNR is included.}
\end{deluxetable}

\begin{deluxetable}{lcc}
\tablewidth{0pc}
\tabletypesize{\scriptsize}
\tablenum{5}
\tablecaption{High Ion Column Density Ratios\tablenotemark{a}} 
\tablehead{
\colhead{Star}&\colhead{N(Si IV)$\lambda$1402/N(O VI)} &
\colhead{N(C IV)$\lambda$1550/N(O VI)}
}
\startdata
AV229                 & $0.27-0.30$  & $0.62-0.70$  \\
AV232                 & $0.19-0.25$  & $0.25-0.37$  \\
AV235                 & $0.08-0.20$  & $\ge0.15$     \\ 
Sk~188                & $0.13-0.23$  & $0.34-0.57$  \\ 
\enddata
\tablenotetext{a}{Column densities were calculated from the equivalent
widths in Fitzpatrick \& Savage 1985, except O VI column densities
from this paper. The column densities for AV229 were remeasured from
the data used by Koenigsberger et al. 2001.}
\end{deluxetable}

\end{document}